\documentclass[10pt]{article}
\usepackage{multirow}
\usepackage[dvips]{graphicx}

\usepackage[psamsfonts]{amssymb}
\usepackage{amsxtra}
\usepackage{threeparttable}
\usepackage{amsmath}
\usepackage{amssymb}

\usepackage{graphicx}

\usepackage{cite}

\usepackage{color}

\usepackage{setspace}
\usepackage{float}

\usepackage{url}

\usepackage[labelfont=bf,labelsep=period,justification=raggedright]{caption}

\makeatletter
\renewcommand{\@biblabel}[1]{\quad#1.}
\makeatother

\date{}

\begin{document}

\begin{flushleft}
{\Large
\textbf{Vibrotactile Stimulus Frequency Optimization for the Haptic BCI Prototype}
}

Hiromu Mori$^1$, 
Yoshihiro Matsumito$^1$, 
Shoji Makino$^1$, 
Victor Kryssanov$^3$, 
\\and Tomasz M. Rutkowski$^{1,2,\ast}$
\\
\bf{1} TARA Life Science Center, University of Tsukuba, Tsukuba, Japan\\
\bf{2} RIKEN Brain Science Institute, Wako-shi, Japan\\
\bf{3} Ritsumeikan University, Kyoto, Japan\\
$\ast$ E-mail: tomek@tara.tsukuba.ac.jp
\end{flushleft}

\section*{Abstract}

The paper presents results from a psychophysical study conducted to optimize vibrotactile stimuli delivered to subject finger tips in order to evoke the somatosensory responses to be utilized next in a haptic brain computer interface (hBCI) paradigm. We also present the preliminary EEG evoked responses for the chosen stimulating frequency. The obtained results confirm our hypothesis that the hBCI paradigm concept is valid and it will allow for rapid stimuli presentation in order to improve information-transfer-rate (ITR) of the BCI.

\section{Introduction}
% no \IEEEPARstart

The state of the art brain computer/machine (BCI/BMI) interfaces relay on visual and imagery paradigms~\cite{bciBOOKwolpaw}, which require longer training or good vision from the subjects. Alternative solutions propose to utilize auditory~\cite{tomekHAID2011} or haptic (somatosensory) modalities~\cite{sssrBCI2006,tomek2012haptic} in order to enhance brain-computer interfacing comfort or to boost information transfer rate (ITR).

A concept of utilizing brain somatosensory (haptic) modality creates a very interesting possibility to target tactile sensory domain, which is not as demanding as vision during operation of machinery or visual computer applications. A potential haptic/somatosensory BCI/BMI paradigm is thus potentially a less mentally demanding. The first successful trial to utilize steady-state somatosensory responses (SSSR) to create the BCI/BMI~\cite{sssrBCI2006} targeted a very low stimulus frequency range of $20-31$Hz to elucidate subject's attentional modulation. The lower frequencies excite Meissner-endings of human finger tips~\cite{natureHAPTIC2009}. We propose to utilize also higher frequencies in a range of $300-400$Hz to stimulate Pacini-endigs~\cite{natureHAPTIC2009}, since these frequencies are very suitable for shorter stimulus delivery which would fit perfectly a somatosensory evoked potential (SEP) generation mechanism in the brain. 

\begin{figure}[H]
	\centering
	\includegraphics[width=0.5\linewidth]{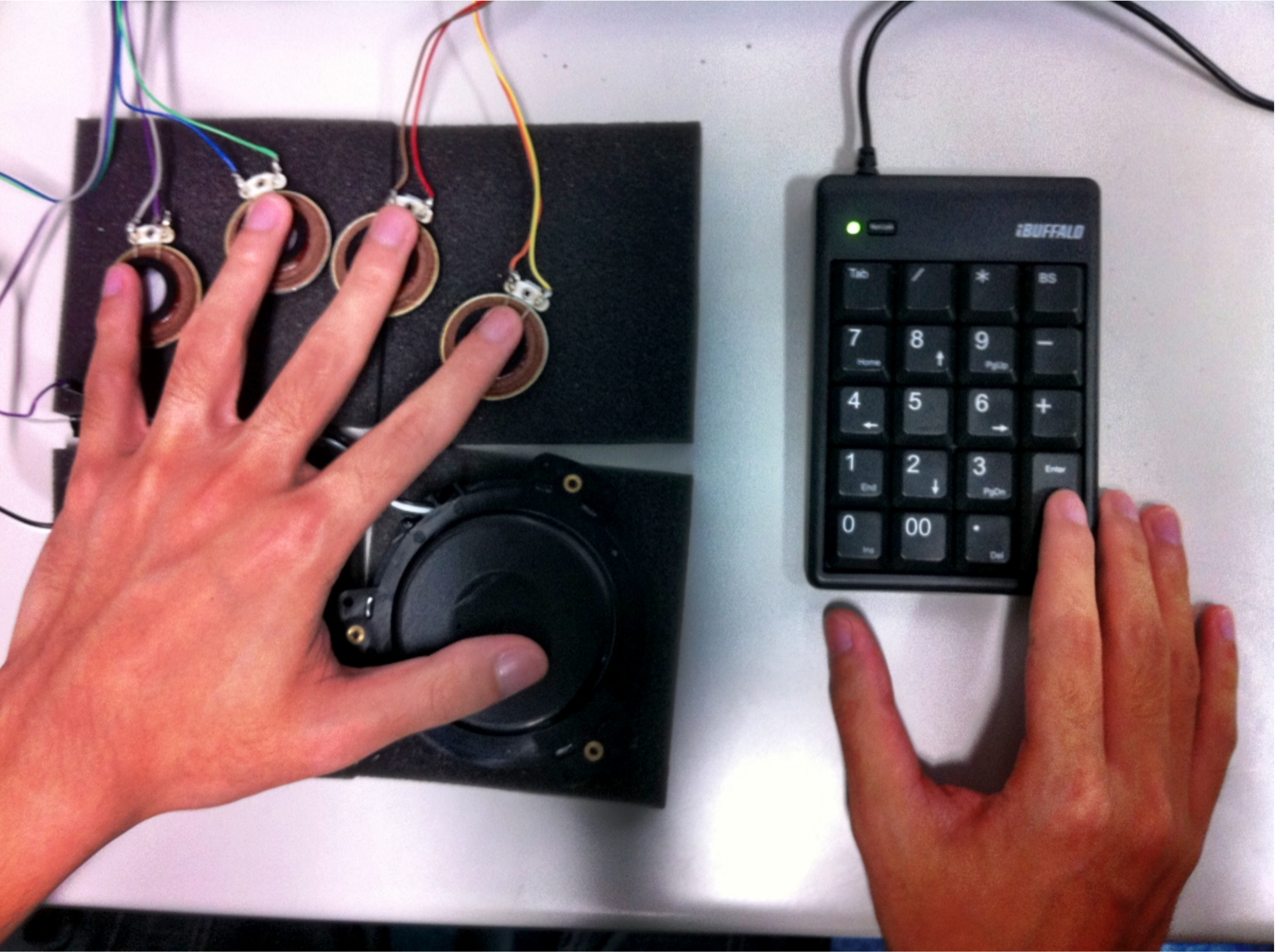}
	\caption{Vibrotactile stimuli delivered to the left hand finger tips and the palm area using acoustic frequency tactile exciters. The right (dominant) hand is positioned on a response pad used in the psychophysical experiment.}\label{fig:hand}
\end{figure}

In order to identify user's preferences before a final BCI/BMI paradigm design which will utilize stimuli--driven responses captured in EEG, we conduct first psychophysical experiments to identify the preferred stimuli frequencies from behavioral responses. This paper reports psychophysical results obtained from stimulation of the five fingers and a palm area of a single hand with various frequencies as depicted in Figure~\ref{fig:hand}. Next we also present the preliminary results with SEP EEG averaged responses to validate our hypothesis of the hBCI concept. 

In the following sections we introduce vibrotactile stimulus delivery technique. Next we describe the psychophysical and EEG experiments. A statistical analysis of the obtained results discussion concludes the paper.

%\subsection{Subsection Heading Here}
%Subsection text here.

%\subsubsection{Subsubsection Heading Here}
%Subsubsection text here.

\section{Methods}

The SEP generated by the vibrotactile stimuli itself recorded in brain waves (EEG) is an evoked transient neural potential caused by the complex tactile stimulus. It is evoked by turning ON/OFF of a tactile vibration impulse~\cite{tomekHAID2011,tomek2012haptic}.

In order to design comfortable and easy to detect by human subjects the vibrotactile stimuli, which next will be utilized in BCI/BMI experiments, we conduct first preliminary experiments in order to determine stimulus difficulty from behavioral responses - ``button presses'' after identified targets. It is known, that a variability of psychophysical response time delays is related to task difficulty and the cognitive load. We aim to determine six vibrotactile stimuli frequencies with uniform task difficulties which could be next plugged into a haptic BCI/BMI paradigm. Next we also test brain SEP response in EEG.

\subsection{Vibrotactile stimuli delivery}

The vibrotactile stimuli were delivered as sinusoidal waves generated by a portable computer with \textsf{Max/MSP} software. The stimuli were generated via six (for psychophysical experiment) and two (for EEG measurement) channel outputs of an external \emph{digital--to--analog} signal converter \textsf{MOTU UltraLite-mk3 Hybrid} coupled with the \textsc{YAMAHA P4050} power amplifiers. The stimuli were delivered to subject fingers via two types of tactile exciters working in the ranges of $300-20,000$~Hz (four small \textsf{HIAX19C01-8}) and $20-80$~Hz (two large \textsf{TT25-16 PUCK}) as depicted in Figure~\ref{fig:hand}. The subjects placed their palms and fingers on the exciters and attended (button--press responded in case of psychophysical or mentally counted only in case of EEG experiments) only to the instructed locations, while ignoring the other stimuli.

\subsection{The Psychophysical Experiment Protocol}

The psychophysical experiments were conducted in order to investigate stimulus carrier frequency relation on subject response time and accuracy.
The behavioral responses were collected via a small numeric keypad as depicted in Figure~\ref{fig:hand}. The subject used their dominant hand for responses, while the other hand was stimulated vibrotactually with various frequencies. 
Each trial was composed of a random order $500$~ms stimulation delivered to each finger with \emph{inter--stimuli--interval} (ISI) of $500$~ms. The subject was instructed in each trial to attend to one stimuli, while ignoring the others.
The response times were registered with the same \textsf{Max/MSP} patch which was used for the stimulus generation.

\subsection{The EEG Experiment Protocol}

The EEG experiments were conducted in order to confirm the SEP responses usability for the BCI/BMI paradigm. The experimental settings were as follows.

The EEG signals were captured with eight dry electrodes portable wireless system by \textsf{g.tec} (\textsf{g.SAHARA \& g.MOBllab+}). The electrodes were attached to following scalp locations \emph{Cz, CPz, POz, Pz, P1, P2, C3,} and \emph{C4.} The recorded EEG signals were processed by \textsf{BCI2000} application~\cite{bci2000book}. The sampling rate was set to $256$Hz, high pass filter at $0.1$Hz, the low pass filter at $40$Hz, and notch filter at $50$Hz.

The two stimuli types with the lengths of $[50, 200]$~ms, or $[50, 150]$~ms, were delivered randomly to the forefingers of subjects' both hands. Both of the stimuli types had the same carrier frequency of $333$~Hz. We chose this frequency since the psychophysical tests (see the summary Figure~\ref{fig:buttonPRESSgm}) suggested that it caused the larger response time variability. During the experiment, each trial was composed of the single target and three non--targets stimuli presentation in a random train repeated ten times. The subject was instructed to attended to target stimulus while ignoring the other three types. This procedure was repeated for each stimulus as the target. The offline BCI paradigm experiments were conducted in accordance with institutional ethical committee guidelines for experiments with human subjects.

\section{Results and Conclusions}

The behavioral results obtained for all subjects separately are presented in Figures~\ref{fig:buttonPRESSs1}-\ref{fig:buttonPRESSs3}, in form of boxplots of the response time distributions. The results analysis have shown that the vibrotactile stimuli frequency does not influence significantly the response delays. We conducted pairwise \emph{t-test} and \emph{Wilcoxon} analyses to confirm the observations. In both cases the means have been confirmed to be similar. The results rejected a hypothesis of mean differences for various vibrotactile stimulating frequencies. A grand mean average for all subjects has been also presented in a summary Figure~\ref{fig:buttonPRESSgm}.

We conclude the psychophysical experiments with an observation that vibrotaclite carrier frequencies in the ranges of $30-50$~Hz and $300-400$~Hz did not influence the instructed vibrotactile target perception and cognition since the behavioral responses were the same, or differed non-significantly. Only the $333$~Hz carrier frequency has caused larger response variability and it was chosen for the subsequent tests in EEG experiments.  

\begin{figure}[H]
	\centering
	\includegraphics[width=0.6\linewidth]{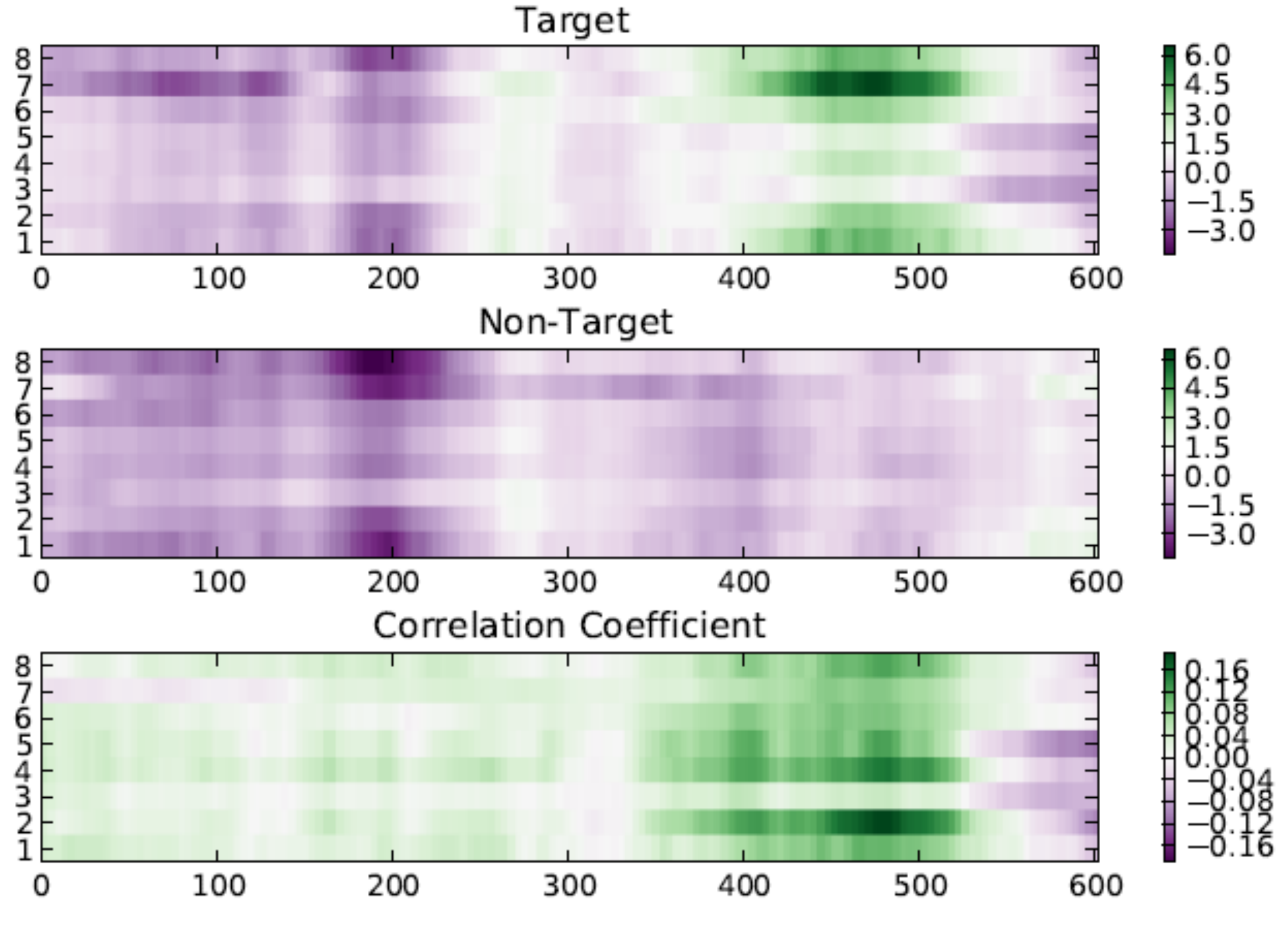}
	\caption{Averaged EEG ERP responses from the subject~$\#1$ for targets (the top panels), non-targets (the middle panels) and in form of correlation coefficients between the above conditions (the bottom panels). The $X-axess$ represent time in milliseconds after the stimuli onset and the $Y-axes$ EEG channels.  The figure was created with \textsf{Py3GUI}~\cite{p3gui}.}
	\label{fig:erpS1}
\end{figure}

\begin{figure}[H]
	\centering
	\includegraphics[width=0.6\linewidth]{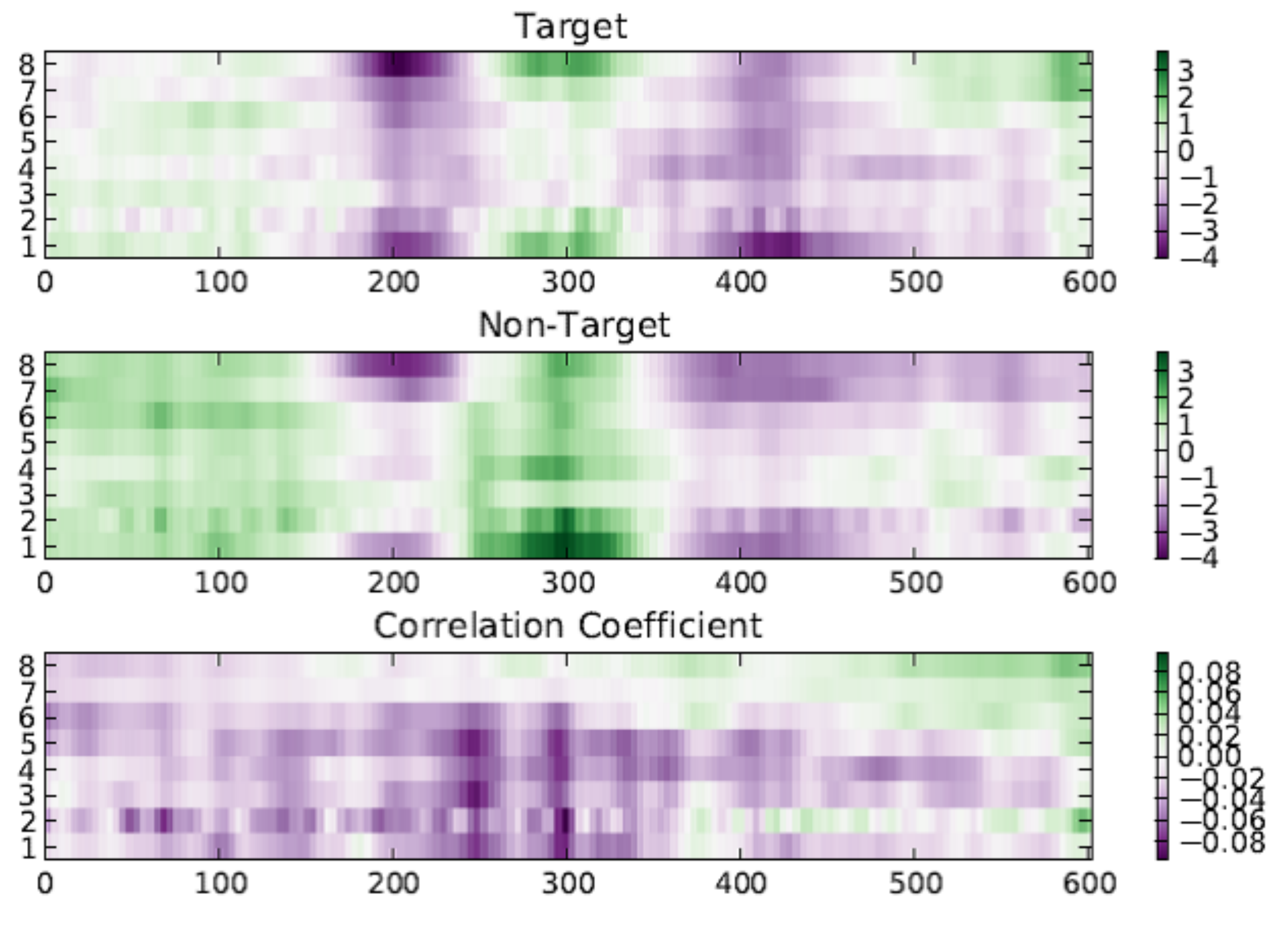}
	\caption{Averaged EEG ERP responses from the subject~$\#2$ for targets (the top panels), non-targets (the middle panels) and in form of correlation coefficients between the above conditions (the bottom panels). The $X-axess$ represent time in milliseconds after the stimuli onset and the $Y-axes$ EEG channels.  The figure was created with \textsf{Py3GUI}~\cite{p3gui}.}\label{fig:erpS2}
	\end{figure}

\begin{figure}[H]
	\centering
	\includegraphics[width=0.6\linewidth]{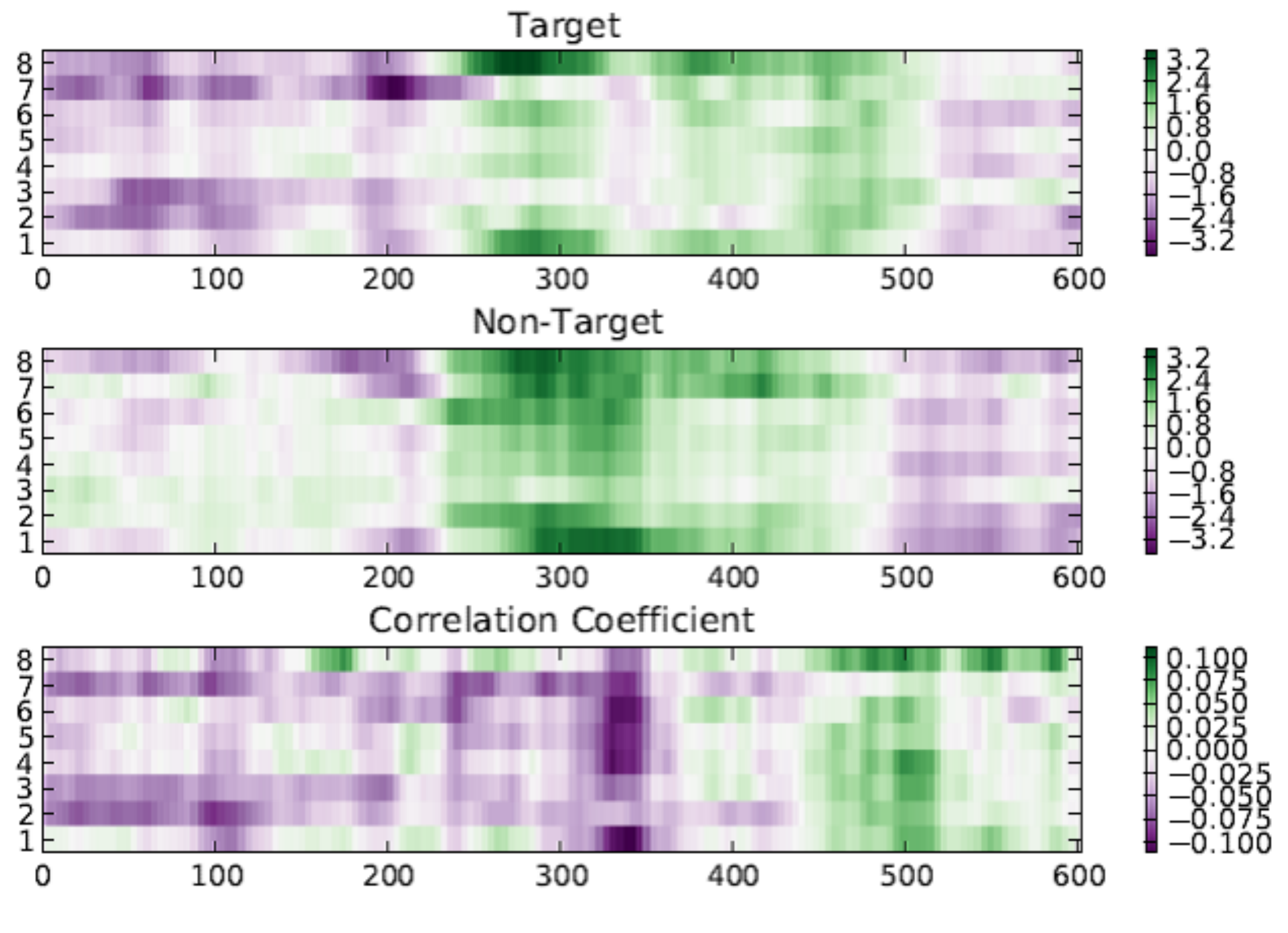}
	\caption{Averaged EEG ERP responses from the subject~$\#3$ for targets (the top panels), non-targets (the middle panels) and in form of correlation coefficients between the above conditions (the bottom panels). The $X-axess$ represent time in milliseconds after the stimuli onset and the $Y-axes$ EEG channels.  The figure was created with \textsf{Py3GUI}~\cite{p3gui}.}\label{fig:erpS3}
\end{figure}

\begin{figure}[H]
	\centering
	\includegraphics[width=.6\linewidth]{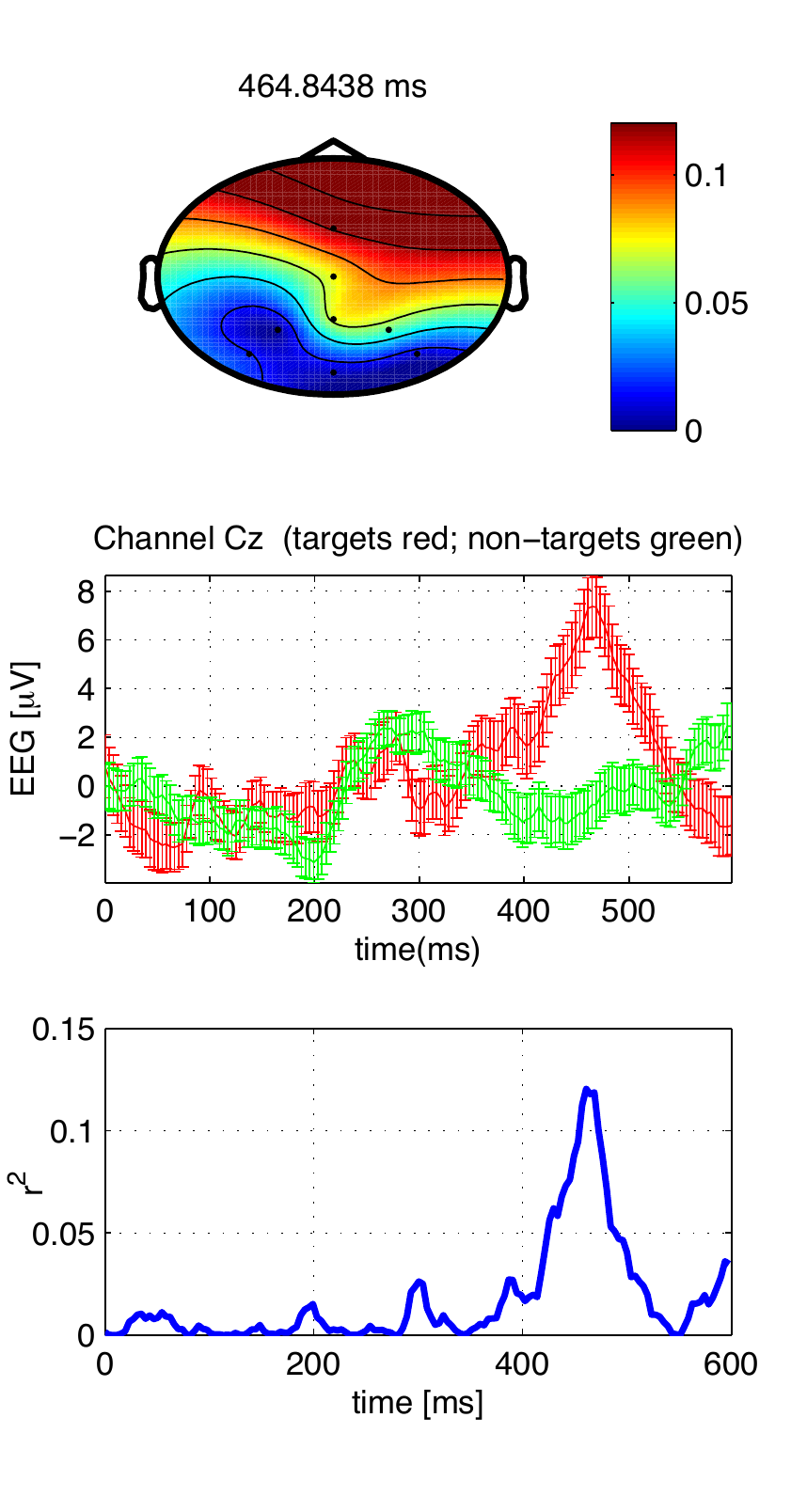}
	\caption{The top panel presents a topographic plot of the single subject SEP response at $464.8438$~ms where the largest difference between target and non-target ERP occurs, as presented in the middle panel together with stander deviation error-bars visualizing the statistical significance (around $400-500$~ms range). The bottom panel presents the signed statistical difference (i.e., signed $r^2$) value which evaluates the discriminability between the two types of ERP. }
	\label{fig:erpFINAL}
\end{figure}

The EEG experiment results has been presented in Figures~\ref{fig:erpS1}--\ref{fig:erpS3} where evoked responses to targets (top panels); non-targets (middle panels); and distance measures in form of correlation coefficients (bottom panels) indicating the features separability. For the three subjects we have observed clear differences in evoked potentials patterns between targets and non-targets, what is also obvious in correlation coefficient diagrams. Each subject on the other hands had various latencies of the strongest ``aha-response'' peaks ranging from around $300-400$~ms. We have presented also in the Figure~\ref{fig:erpFINAL} the results for a single subject in form of evoked response head topography and a single channel averaged SEP for targets and non-targets together with standard deviation error bars, which identified the statistically significant difference around $400-500$~ms for this particular subject. The signed statistical difference $r^2$ graph further confirmed the SEP separability in the above latency range.

In Table~\ref{tab:results} we also have presented the preliminary BCI experiment classification results with utilization of the \textsf{P300Classifier}~\cite{p300classifier} package with ITR calculated in form of \emph{bit--per--minute--rate} (BPMR) as proposed in~\cite{bciSPATIALaudio2010}. The results for the three subjects are very promising taking into account a very conservative ISI of $600$~ms which in our future planned experiments will be significantly shortened.  

As the final conclusion of the presented experiments we report on the very promising vibrotactile (haptic) BCI/BMI modality which is a step forward in creation of the more friendly paradigms.

\section*{Acknowledgments}

This research was supported in part by the Strategic Information and Communications R\&D Promotion Programme no. 121803027 of The Ministry of Internal Affairs and Communication in Japan, and by KAKENHI, the Japan Society for the Promotion of Science grant no. 12010738. We also acknowledge the technical support from YAMAHA Sound \& IT Development Division in Hamamatsu, Japan.

\bibliography{hbci}

\newpage
\section*{Figure Legends}

\begin{figure}[H]
	\centering
	\includegraphics[width=0.65\linewidth]{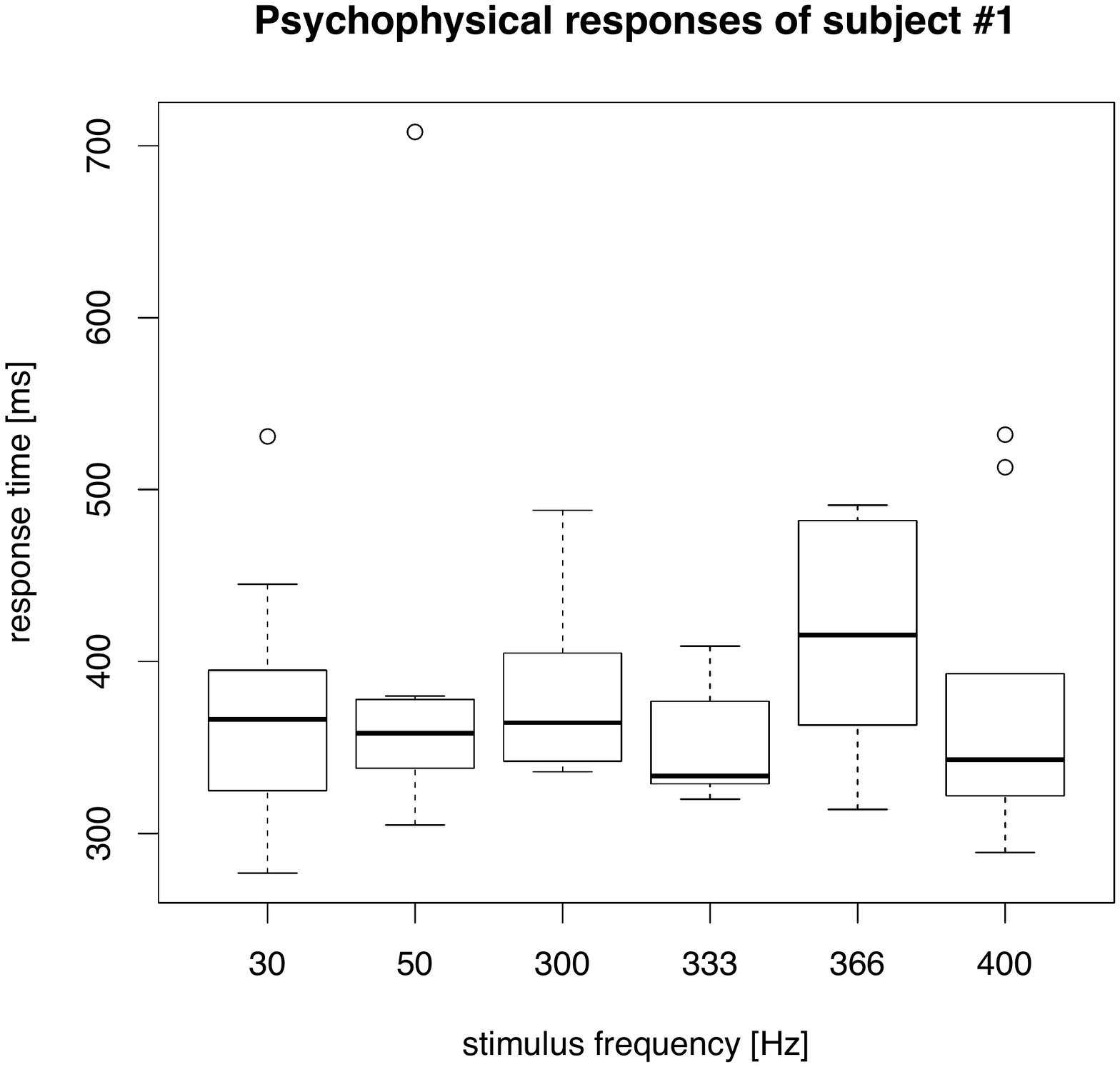}
	\caption{Psychophysical experiment results in form of response time delays to various vibrotactile stimulus frequencies for subject $\#1$. The differences among stimulus frequencies are not significant in the tested frequency ranges.}\label{fig:buttonPRESSs1}
\end{figure}

\begin{figure}[H]
	\centering
	\includegraphics[width=0.65\linewidth]{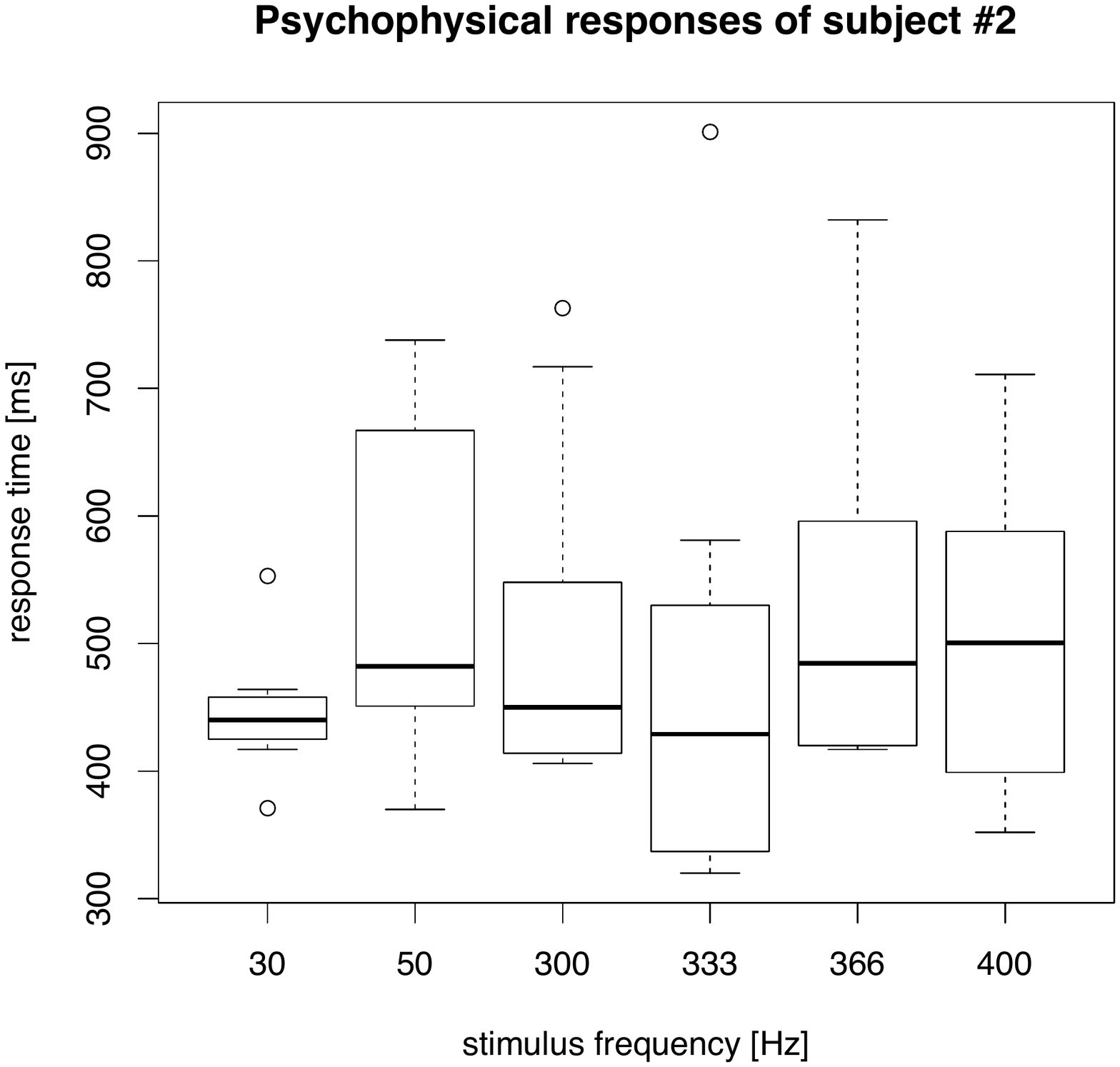}
	\caption{Psychophysical experiment results in form of response time delays to various vibrotactile stimulus frequencies for subject $\#2$. The differences among stimulus frequencies are not significant in the tested frequency ranges.}\label{fig:buttonPRESSs2}
\end{figure}

\begin{figure}[H]
	\centering
	\includegraphics[width=0.65\linewidth]{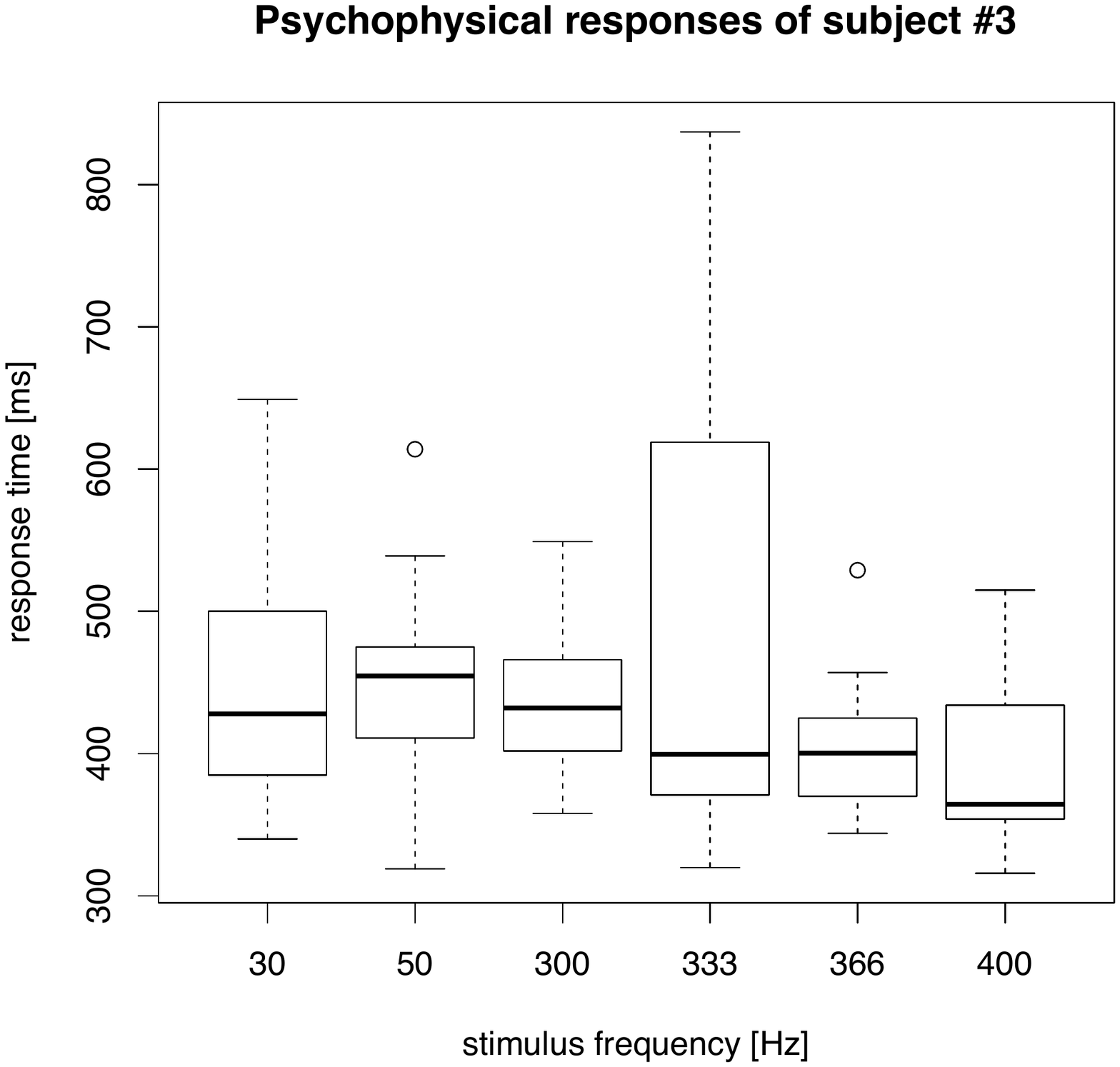}
	\caption{Psychophysical experiment results in form of response time delays to various vibrotactile stimulus frequencies for subject $\#3$. The differences among stimulus frequencies are not significant in the tested frequency ranges.}\label{fig:buttonPRESSs3}
\end{figure}

\begin{figure}[H]
	\centering
	\includegraphics[width=0.5\linewidth]{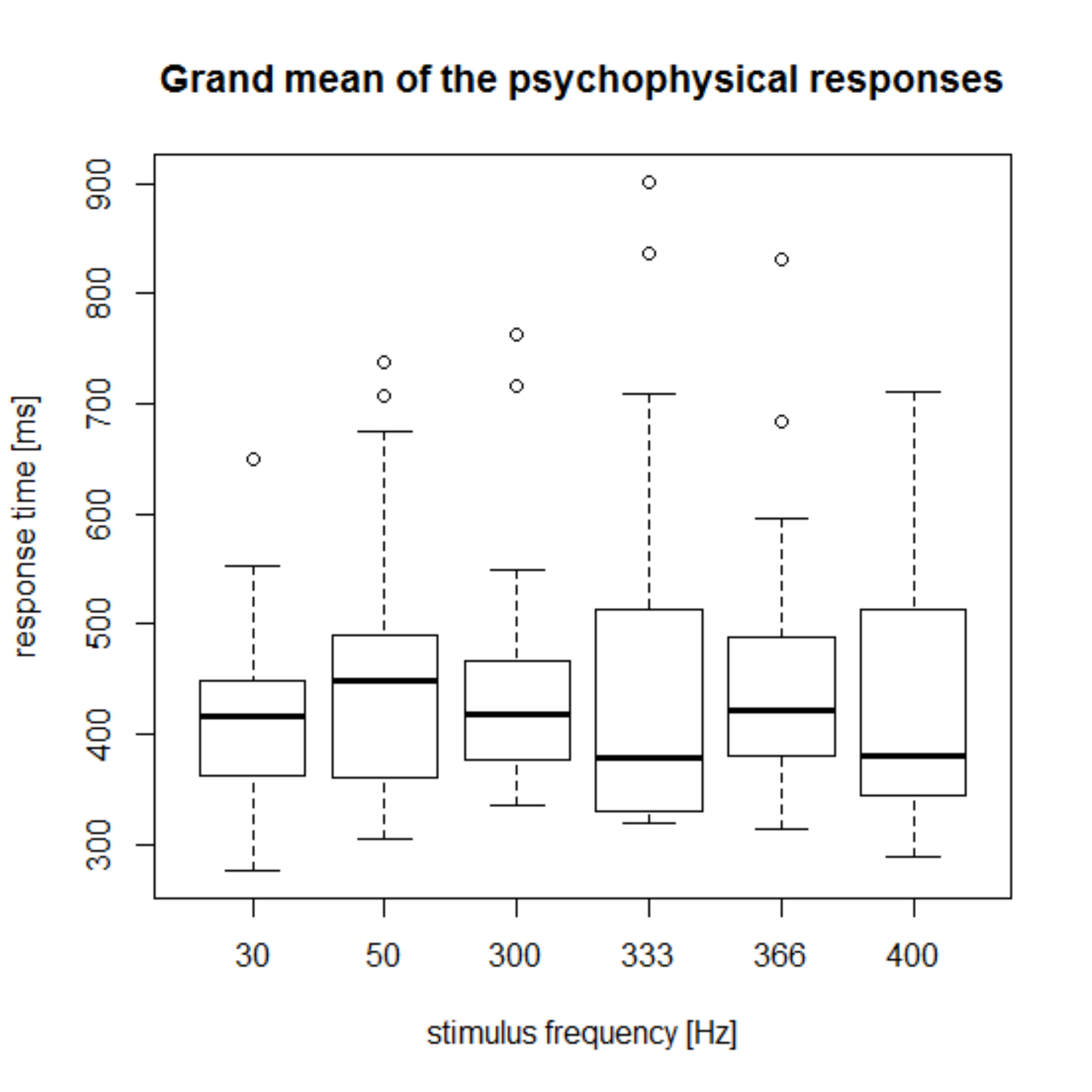}
	\caption{Grand mean average of the psychophysical response results for all subjects as from the experiments depicted in Figures~\ref{fig:buttonPRESSs1}--	\ref{fig:buttonPRESSs3}.}
	\label{fig:buttonPRESSgm}
\end{figure}

\begin{description}
	\item[Figure~\ref{fig:hand}] Vibrotactile stimuli delivered to the left hand finger tips and the palm area using acoustic frequency tactile exciters. The right (dominant) hand is positioned on a response pad used in the psychophysical experiment.
	\item[Figure~\ref{fig:buttonPRESSs1}] Psychophysical experiment results in form of response time delays to various vibrotactile stimulus frequencies for subject $\#1$. The differences among stimulus frequencies are not significant in the tested frequency ranges.
	\item[Figure~\ref{fig:buttonPRESSs2}] Psychophysical experiment results in form of response time delays to various vibrotactile stimulus frequencies for subject $\#2$. The differences among stimulus frequencies are not significant in the tested frequency ranges.
	\item[Figure~\ref{fig:buttonPRESSs3}] Psychophysical experiment results in form of response time delays to various vibrotactile stimulus frequencies for subject $\#3$. The differences among stimulus frequencies are not significant in the tested frequency ranges.
	\item[Figure~\ref{fig:buttonPRESSgm}] Grand mean average of the psychophysical response results for all subjects as from the experiments depicted in Figures~\ref{fig:buttonPRESSs1}--	\ref{fig:buttonPRESSs3}.
	\item[Figure~\ref{fig:erpS1}] Averaged EEG ERP responses from the subject~$\#1$ for targets (the top panels), non-targets (the middle panels) and in form of correlation coefficients between the above conditions (the bottom panels). The $X-axess$ represent time in milliseconds after the stimuli onset and the $Y-axes$ EEG channels.  The figure was created with \textsf{Py3GUI}~\cite{p3gui}.
	\item[Figure~\ref{fig:erpS2}] Averaged EEG ERP responses from the subject~$\#2$ for targets (the top panels), non-targets (the middle panels) and in form of correlation coefficients between the above conditions (the bottom panels). The $X-axess$ represent time in milliseconds after the stimuli onset and the $Y-axes$ EEG channels.  The figure was created with \textsf{Py3GUI}~\cite{p3gui}.
	\item[Figure~\ref{fig:erpS3}] Averaged EEG ERP responses from the subject~$\#3$ for targets (the top panels), non-targets (the middle panels) and in form of correlation coefficients between the above conditions (the bottom panels). The $X-axess$ represent time in milliseconds after the stimuli onset and the $Y-axes$ EEG channels.  The figure was created with \textsf{Py3GUI}~\cite{p3gui}.
	\item[Figure~\ref{fig:erpFINAL}] The top panel presents a topographic plot of the single subject SEP response at $464.8438$~ms where the largest difference between target and non-target ERP occurs, as presented in the middle panel together with stander deviation error-bars visualizing the statistical significance (around $400-500$~ms range). The bottom panel presents the signed statistical difference (i.e., signed $r^2$) value which evaluates the discriminability between the two types of ERP.
\end{description}

\newpage
\section*{Tables}

\begin{table}[H]
\renewcommand{\arraystretch}{1.3}
\caption{BCI EEG experiment classification accuracy and bit--per--minute--rate results.}
\label{tab:results}
\centering
\begin{tabular}{|c|c|c|c|}
\hline
Subject & Number of averages & The best accuracy & BPMR \\
\hline \hline
$\#1$ & $1$ & $75\%$ & $19.81$ bit/min \\
\hline
$\#2$ & $2$ & $50\%$ & $2.59$ bit/min \\
\hline
$\#3$ & $1$ & $75\%$ & $19.81$ bit/min \\
\hline
\end{tabular}
\end{table}

\newpage
\section*{Figures}

\end{document}